# Intrusions and turbulent mixing above a small Eastern Mediterranean seafloor-slope

by Hans van Haren

Royal Netherlands Institute for Sea Research (NIOZ), P.O. Box 59, 1790 AB Den Burg, the Netherlands.
e-mail: hans.van.haren@nioz.nl

ABSTRACT: Growing evidence is found in observations and numerical modelling of the importance of steep seafloor topography for turbulent diapycnal mixing leading to redistribution of suspended matter and nutrients, especially in waters with abundant internal tides. One of the remaining questions is the extent of turbulent mixing away from and above nearly flat topography, which is addressed in this paper. Evaluated are observations from an opportunistic, week-long mooring of high-resolution temperature sensors above a small seafloor slope in about 1200 m water depth of the Eastern Mediterranean. The environment has weak tides, so that near-inertial motions and -shear dominate internal waves. Vertical displacement shapes suggest instabilities to represent locally generated turbulent overturns, rather than partial salinity-compensated intrusions dispersed isopycnally from turbulence near the slope. This conclusion is supported by the duration of instabilities, as all individual overturns last shorter than the mean buoyancy period and sequences of overturns last shorter than the local inertial period. The displacement shapes are more erratic than observed in stronger stratified waters in which shear drives turbulence, and better correspond with predominantly buoyancy-driven convection-turbulence. This convection-turbulence is confirmed from spectral information, generally occurring dominant close to the seafloor and only in weakly stratified layers well above it. Mean turbulence values are 10-100 times smaller than found above steep ocean topography, but 10 times larger than found in the open-ocean interior.

# 1. Introduction

Turbulent mixing in the deep ocean is generally a slow process, in comparison with violent crashing of surface waves breaking at a beach. Here, 'diapycnal' turbulent mixing is meant across the stable vertical density stratification (Garrett 1982). This mixing is sometimes called 'small-scale' turbulent mixing or 'irreversible' turbulent mixing, although the latter adjective is somewhat redundant as turbulent mixing is an irreversible process in the commonly modelled downgradient transfer of momentum to heat which dissipates the mechanical energy (Tennekes and Lumley 1972). An energy source is thus required to maintain turbulence. With the transfer, turbulence can mix quantities like oxygen, suspended matter, and nutrients. In bulk parametrization models, the diapycnal turbulent flux is represented by a common diffusion coefficient $K_z$ to be multiplied by the vertical gradient of any quantity.

The naming 'small-scale' may come from the need to avoid confusion and make a distinction with larger scale 'horizontal dispersion' by 'mesoscale variability', commonly known as 'mesoscale eddies'. To fit in general ocean circulation models this dispersion is parameterized in a similar manner by using horizontal diffusivity coefficient $K_h$. While important for transport especially near western boundary flows, mesoscale eddy dispersion is not a physical turbulence process, as only little of it irreversibly transfers momentum to heat (Wunsch and Ferrari 2004). It bears similarities with important horizontal dispersion by tidal currents in shallow seas (Zimmerman 1986). Mesoscale eddy fluxes are at least partially modelled upgradient so that $K_h$ becomes negative (Kamenkovich et al. 2021). Also, mesoscale dispersion will not reach isotropy in three spatial dimensions because the horizontal scales are 3-4 orders of magnitude larger than the vertical one. Due to this large difference in scales, it cannot be called '2D-turbulence', because fully developed turbulence is 3D and isotropic by nature. The latter is named 'real turbulence' by Fjørtoft (1953) to distinguish from 2D vorticity-preserving, besides kinetic energy-preserving, flows. Here we drop the word 'real' for turbulence.

Some of the above confusion may come from interpreting a statement by Wunsch (1981) that the entire ocean is turbulent. This cannot be true, as turbulence is a flow property. What is meant is that virtually everywhere in the ocean the water-flow is turbulent, even at speeds of 0.01 m s$^{-1}$ in the deep-ocean: The bulk Reynolds number is typically $O(10^5$-$10^7)$, and thus very high so that "the ocean appears to be in a turbulent state" (Wunsch 1981). Thus, while mesoscale eddies, or more generally low-frequency flow variability with timescales longer than one day, are important for the advective transport of suspended matter and heat, their individual flows may be turbulent, but their shape and structure are not.

In a similar fashion as mesoscale dispersion, but on much smaller scales, the generally stably stratified ocean is not considered to be fully turbulent. This is because, assuming weak horizontal density gradients, a considerable inhibition, albeit not blocking!, exists of vertical, diapycnal exchange in comparison with horizontal components (Thorpe 2005). Only at scales smaller than the Ozmidov (1965)-scale O(1-10 m) is isotropy expected, allowing full 3D developed turbulence. Isotropy of relatively large scale is thus expected in the largest near-homogeneous layers, which are created mainly by internal wave straining in the ocean interior, and mainly by internal wave breaking over steep seafloor topography. From an active turbulence patch, whether created in the interior or near the seafloor, the decaying, irreversibly dissipative turbulence is thought to be more easily dispersed along than across isopycnals in the interior (Thorpe 2005).

Debate is ongoing as to how far such isopycnal dispersal is carried from the turbulence patch. To which one can add whether such isopycnal dispersal, or 'intrusion', being governed by pressure gradients and shear-flow, may also not overturn by itself thereby locally generating turbulence. Or, whether such small-scale horizontal dispersion is more akin to large-scale dispersion by mesoscale eddies, and, for example, whether a horizontal diffusivity may apply to intrusions. It has been shown that previous estimates of horizontal dispersal distances O(100 km), as suggested and sketched by e.g. Armi (1979), are too large by at least a factor of 10 for isopycnal dispersal of turbulence generated by internal wave breaking above steep topography (Winters 2015; van Haren 2023a). Another question arises, that while above sloping topography an active turbulent homogenizing patch may disperse at its interior isopycnal level, one wonders what occurs with dispersion from the same homogenized patch at interior non-isopynal levels just above and below? Modeling results by Winters (2015) demonstrate dispersal in various isopycnal layers stacked on top of each other, presumably maintained by locally active turbulence consisting of static instabilities whilst being driven by momentum fed by the active turbulence at the steeply sloping seafloor.

In this paper, we are interested in the apparent development of intrusions or turbulent layering under conditions of the Eastern Mediterranean Sea where: a) the local seafloor slope is small instead of steep, b) local turbulence generation is not tide-driven and, because of the small slope, at least two orders of magnitude weaker than above steep seamounts in, e.g., the Atlantic Ocean. The lack of any substantial internal tide bears similarity with arctic seas, poleward from a latitude of 74° at which freely propagating semidiurnal lunar internal tidal waves reflect equatorward. In such seas around Spitsbergen, relatively strong turbulence was observed in microstructure profiler data in apparent partially salinity-compensated intrusions (Padman and Dillon 1991). Turbulence enhancement showed distinct diurnal variability, which was associated with the local tides interacting with topography. However, considering that freely propagating diurnal internal tides certainly cannot exist in the area, the diurnal turbulence variability may be interpreted as the duration of the apparent intrusions found in moored high-resolution temperature (T-)sensor data (van Haren and Greinert 2016). Observations using similar moored T-sensors in the Eastern Mediterranean are discussed here.

**2. Distinguishing turbulence from salinity compensation**

Turbulence works against density differences, which are mainly determined by temperature and salinity variations in the ocean. In the lack of density and/or salinity measurements, one may revert to temperature observations only, provided a tight and consistent density-temperature relationship exists, which implicitly accounts for salinity contributions to density variations.

There are several ways to distinguish turbulent overturns from salinity (over-)compensated intrusions that provide false turbulence values calculated from temperature observations. In stratified waters, turbulent overturning cannot last longer than the longest buoyancy period, which is restricted by the local inertial period in the weakest stratified near-homogeneous layers. Intrusions can last longer than the inertial period because there is no physical restriction near a source, albeit possibly a transfer may develop to vortex motion under the influence of rotation away from a source.

Additionally, for better distinguishing turbulent overturns from false-turbulent ones, the overturn-shapes can be investigated in the instability displacement-vertical (d-z) plane. Instability displacements can be determined following the method by Thorpe (1977) of reordering a vertical density profile containing (turbulence) instabilities in a statically stable one. Van Haren and Gostiaux (2014) provide estimates of the slopes z/d of vertical displacement profiles using models of different idealized overturns. Most observed turbulent overturns do not compare with a purely mechanical solid-body rotation that forms a slope of z/d = ½, but with a half-turn Rankine vortex. The Rankine vortex is a 2D-model of a rotating eddy (vortex) in a viscous fluid. It is used for modelling atmospheric tornadoes. When used in the x-z plane, it can modify density in a stably stratified fluid. After integrating the Rankine motion over half a turn, the disturbed density field creates z/d slopes that are characterized by: a) borders strictly along a slope just z/d > 1 and which are formed by the upper and lower parts of the density overturn portion (i.e., the parts generated by the edge of the vortex), and b) a slope ranging between ½ < z/d < 1 for the forced inner vortex part and mainly denoting the mechanical overturn. Thus, the Rankine vortex introduces two slopes, one which aligns with the slope of a solid-body overturn forming the long inner axis of a parallelogram, and the other which aligns with the maximum displacements possible in an overturn and delineating the sloping edges of a parallelogram. An idealized model of a half-turn Rankine vortex will be presented along with observational data for comparison.

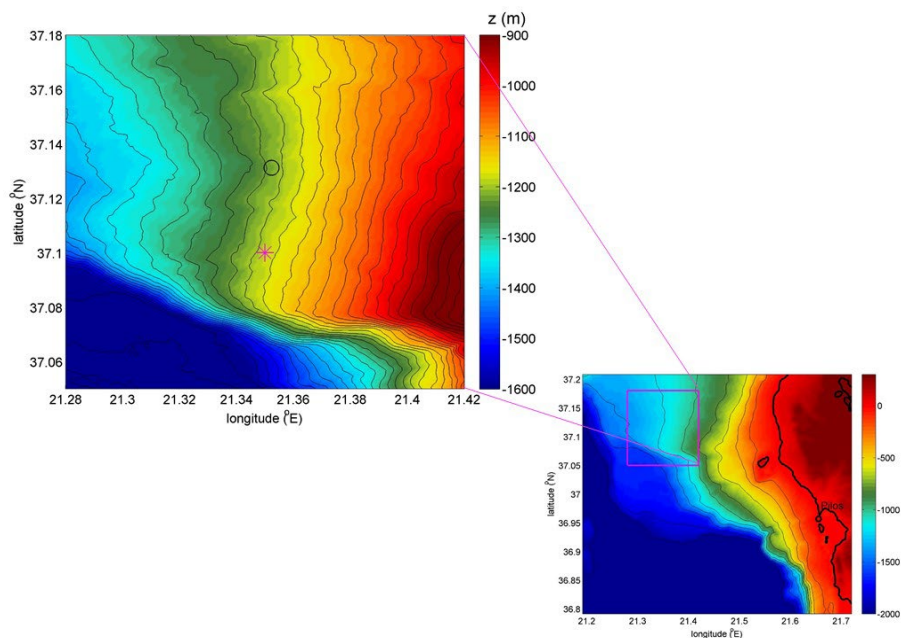

FIG. 1
Small part of Eastern Mediterranean Sea, with black isobaths every 25 m, mooring site (black circle) and yoyo-CTD station (magenta star) offshore of southwestern Peloponnesus (Greece) from 3.75″ EMODnet data. Note the change in colour-range with respect to the larger area plot from 15″ GEBCO23 data with 200-m isobaths and coastline (thick black).

In contrast, a salinity-compensated intrusion of warm-salty water penetrating colder-fresher water of virtually the same density, such as occurring around Mediterranean outflow lenses in the surrounding mid-depth Northeast-Atlantic Ocean, is modeled by a slightly oblique S-shape in density with diffusion/convection incorporated in the intrusive layer (van Haren and Gostiaux 2014). The resulting displacement-shape heavily favors a slope of $z/d \geq 1$ even for the long inner axis of the parallelogram. It thus distinguishes from 1D-solid-body and 2D-Rankine overturns.

**3. Technical details of site and instrumentation**

To study non-semidiurnal lunar tidal internal waves and their induced turbulence dynamics above weakly sloping topography, an 80-m tall taut-wire mooring was in the Eastern Mediterranean Sea just West of Greek Peloponessos at 37°07.87′N, 021°21.14′E, in 1208 m water depth (Fig. 1). For logistic reasons to support various mechanical mooring lowering tests, the 80-m mooring was underwater for nearly 8 days only, between 27 January and 04 February 2011. This period ran from neap to spring of the very weak tide (nearby Pylos-harbour tidal sea level amplitudes vary between 0.05 and 0.15 m for neaps and springs, respectively).

The local seafloor was relatively flat for the testing with a slope of about $\gamma = 0.03$ (1.7°) computed over 6-km horizontal distance (Fig. 1). Under sufficiently density-stratified waters this slope is supercritical for linearly propagating internal waves when their characteristics along which energy is transposed have a slope of $\beta < \gamma$,

$$\beta = \sin^{-1}((\omega^2-f^2)^{1/2}/(N^2-f^2)^{1/2}), \qquad (1)$$

e.g., LeBlond and Mysak (1978). Here, $\omega$ denotes the internal wave frequency, $f = 8.803 \times 10^{-5}$ s$^{-1}$ the local planetary inertial frequency (Coriolis parameter), and $N = 6.6 \times 10^{-4}$ s$^{-1}$ the 100-m large-scale mean local buoyancy frequency. Slope-criticality, $\beta = \gamma$, is reached for $\omega = 9.0 \times 10^{-5}$ s$^{-1} \approx 1.025f$, so that the weak seafloor slope is only supercritical for inertial-frequency internal waves and sub-critical for most others.

The local horizontal Coriolis parameter, which is relevant for internal wave dynamics in weakly stratified waters $N = O(f)$, amounts $f_h = 1.163 \times 10^{-4}$ s$^{-1}$. In such waters, the non-traditional full inertio-gravity wave (IGW) band is not between bounds $\omega \in [f, N]$, for $N \gg f$, but $\omega \in [\omega_{min}<f, \omega_{max}>N]$, for $N \sim O(f)$:

$$[\omega_{min}, \omega_{max}] = (s -/+ (s^2 - f^2N^2)^{1/2})^{1/2}, \qquad (2)$$

using $2s = N^2 + f^2 + f_h^2\cos^2\alpha$, in which $\alpha$ is the angle to the north and $\alpha = 0$ denoting meridional propagation (e.g., LeBlond and Mysak 1978; Gerkema et al. 2008). IGWs can propagate along much steeper characteristics than in traditional (1) and will, partially, experience the present seafloor slope as sub-critical also at f, although waves at the latter frequency will have one (of two) characteristic that is exactly horizontal and which experiences every seafloor slope as super-critical (e.g., Gerkema et al. 2008).

For calibration and temperature-density relationship purposes, a comprehensive set of 36 vertical profiles of shipborne Conductivity Temperature Depth (CTD) were obtained during a period of 12.5 h when the ship was held stationary within a horizontal radius of 10 m. The profiles were made, down to 10 m from seafloor for safety reasons, at 37°06.00′N, 021°21.00′E, 1188 m water depth, between 30 January 16:00 and 31 January 04:24 2011. The CTD-site was 3.5 km due south of the mooring at a similarly weakly sloping seafloor. The site was about 2.5 km north of a tenfold steeper seafloor slope (Fig. 1), which most freely propagating internal waves will experience as super-critical.

The mooring consisted of two acoustic Nortek AquaDopp single-point current meters (CM) 150 m above and directly below a cable holding 61 high-resolution temperature (T) sensors. It was held tautly upright by a single sub-surface 'top'-buoy. The top-buoy provided about 1.7-kN net buoyancy to the entire mooring assembly. Under maximum 0.05 m s$^{-1}$ current amplitudes, the low-drag mooring did not deflect from the vertical by more than 0.7°, i.e., the top-buoy moved <1 m horizontally and <0.01 m vertically, as inferred from tilt and pressure sensors.

The CM sampled at a rate of once per 60 s, while the T-sensors sampled at a rate of once per 1 s. The 'NIOZ4' T-sensors were at 1.0 m intervals in the range of height h = 6-66 m above the seafloor.

NIOZ4 are self-contained T-sensors, with a sensor tip smaller than 1 mm, a response time of <0.5 s, an initial drift of about $1\times10^{-3}$ °C mo$^{-1}$ after aging of the thermistor electronics, a noise level of $<1\times10^{-4}$ °C, and a precision (relative accuracy) better than $5\times10^{-4}$ °C after drift-correction for standard programmed measuring of the Wien Bridge oscillator for the duration of 0.12 s per sample (van Haren 2018). Every 4 h, all sensors are synchronized via induction to a single standard clock, so that the entire 60-m vertical range is sampled in less than 0.02 s.

Of the 61 T-sensors, 1 did not function and 15 showed calibration, noise, or other electronic problems. The data of these sensors are replaced by linear interpolation between neighboring sensors for displaying purposes. Spectra are computed from the original data, after noise correction, except for data from the one malfunctioning sensor. Bias and calibration problems are low-frequency sub-inertial phenomena and do not affect the internal wave – turbulence part of the spectra we are interested in here. However, data from T-sensors showing these electronic problems are excluded from turbulence calculations, as well as their interpolated values as these bias turbulence values low.

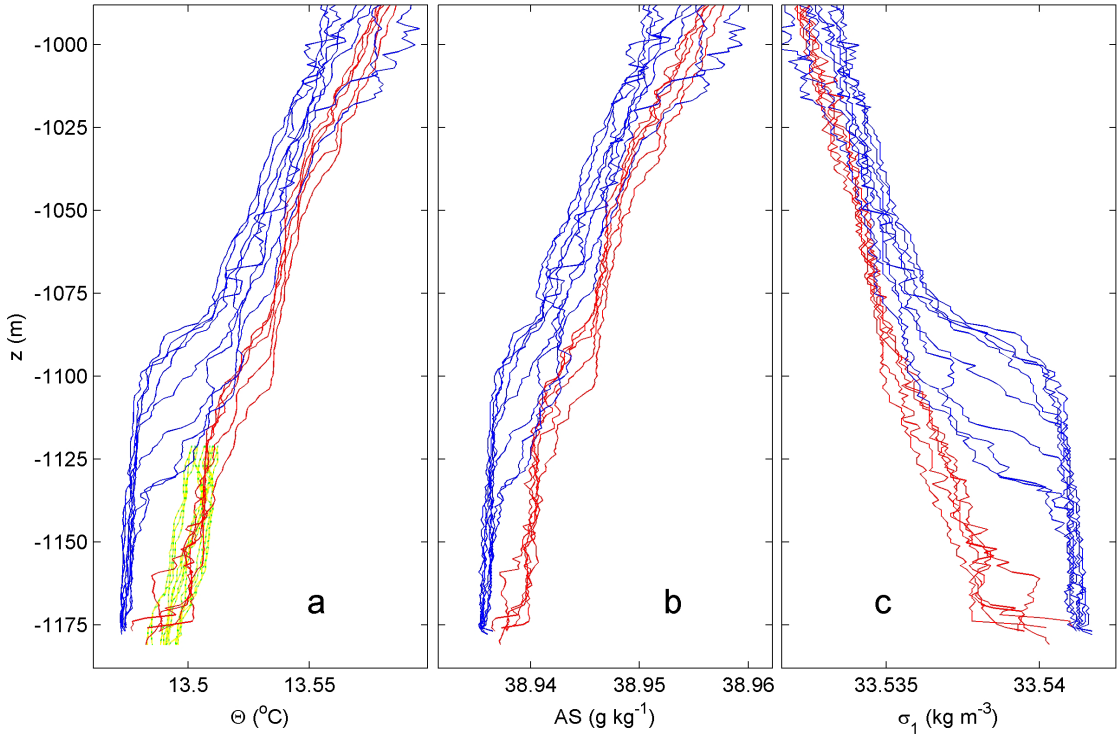

FIG. 2

Shipborne CTD-observations over 200-m vertical range, obtained 3.5 km south of the T-sensor mooring three days after mooring deployment. Plotted are hourly profiles over a 12.5-h period. The local seafloor is at the level of the horizontal axes. (a) Conservative Temperature. Two data groups (blue, red) are distinguished, see text. In yellow-green-dashed the simultaneous moored T-sensor profiles, which are shifted vertically by +20 m to match the local water depths. (b) Absolute Salinity. The x-axis range approximately corresponds with that of a. in terms of density contributions. (c) Density anomaly referenced to 1000 dbar ($10^7$ N m$^{-2}$).

This first-produced batch of NIOZ4 T-sensors showed some high-frequency spikes that were removed during post-processing. The in situ CTD-calibrated moored T-sensor data were bias-corrected to a smooth third-order polynomial mean profile and converted into 'Conservative' (~potential) Temperature data $\Theta$ (IOC, SCOR, IAPSO 2010).

The post-processed moored T-sensor data act as a tracer for variations in density anomaly $\sigma_{1.2}$ following the relation,

$$\delta\sigma_{1.2} = \alpha\delta\Theta, \quad \alpha = -0.125 \pm 0.005 \text{ kg m}^{-3} \text{ °C}^{-1}, \tag{1}$$

where subscript 1.2 indicates a pressure reference of 1200 dbar (1 dbar = $10^4$ N m$^{-2}$). This temperature-density relation is established from data between -1175 < z < -975 m of the yoyo CTD-profiles (Figs 2,3). Given the reasonably tight and consistent relationship (1), the number of moored T-sensors and their spacing of 1.0 m, in combination with their low noise level and precision, allows for accurately calculating values of turbulent kinetic energy dissipation rate $\varepsilon$ and eddy diffusivity $K_z$ via the reordering of unstable overturns thereby making vertical density profiles statically stable as proposed by (Thorpe 1977), see Appendix A. The reordered profiles are used to calculate, small-O(1)-m-scale buoyancy frequency $N_s$.

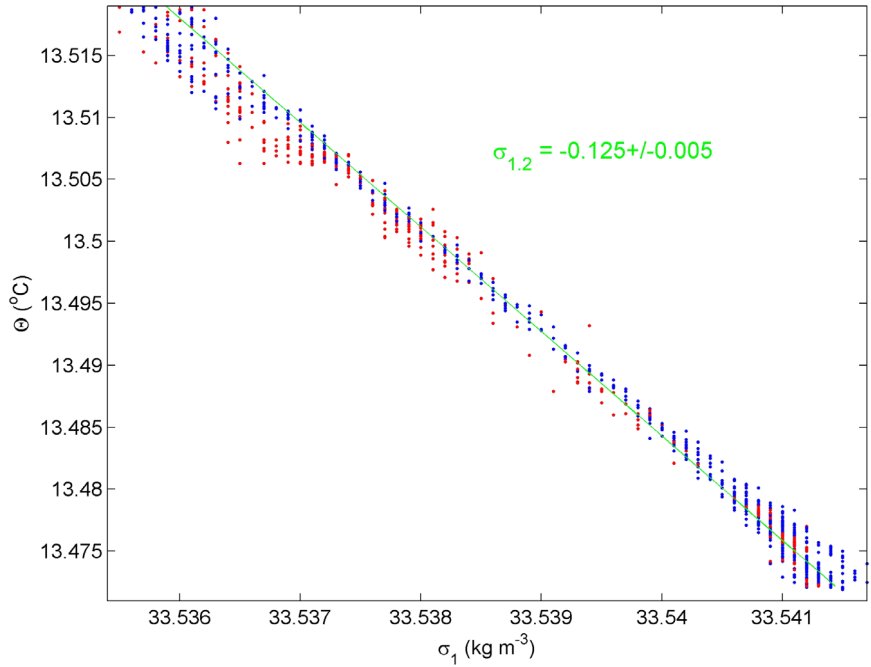

FIG. 3
Conservative Temperature-density anomaly relationship with linear fit in green, from the CTD-profile portions in Fig. 2a and 2c.

## 4. Observations

*a. Overview*

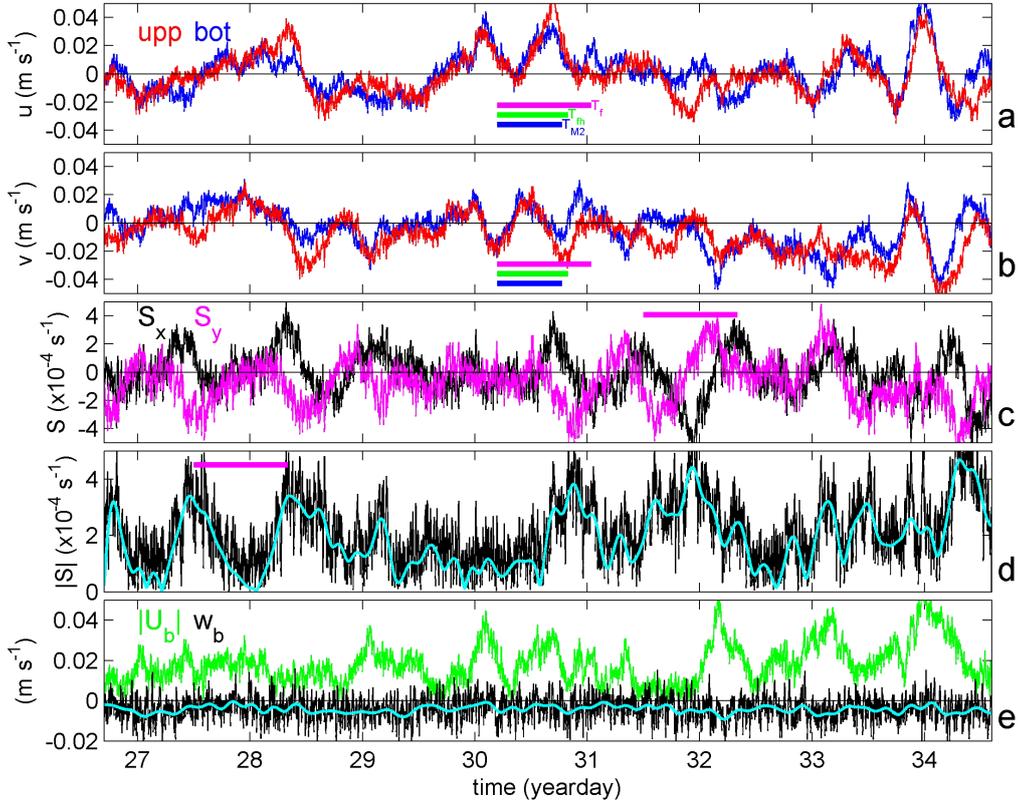

FIG. 4

7.9-Day time series of moored current meter (CM) data. (a). East(positive)-West current component for upper (red) and lower (blue) CM. The horizontal coloured-bars indicate reference times of one: inertial period (magenta), horizontal Coriolis-parameter period (green), semidiurnal lunar period (blue). (b) As a., but for the North(positive)-South current component. (c) Shear in East-West (black) and North-South (magenta) directions computed over the 215-m vertical range between the two CMs. Only one inertial period is indicated. (d) Shear magnitude from c., with low-pass noise-filtered data in cyan. (e) Near-bottom current magnitude from a.,b. (green) in comparison with the local vertical current component (black) and its low-pass noise-filtered version (cyan).

The vertical profiles of shipborne CTD data demonstrate steady decreases of Conservative Temperature (Fig. 2a) and Absolute Salinity (Fig. 2b) with depth. Their contributions of stabilizing temperature stratification and destabilizing salinity stratification result in net stable vertical density profiles (Fig. 2c). For the 12.5-h period of yoyo CTD profiles, of which every third one is shown in Fig. 2, two distinctions can be made in the lower 100 m above the seafloor. Either a relatively strong stratification caps a weakly stratified layer of 30-80 m in thickness (blue profiles), or a moderate stratification reaches to $h < 15$ m and may reside over a thin strongly stratified layer just above the seafloor (red profiles). The simultaneous moored T-sensor profiles uniquely follow the latter, although they do not show a thin layer stratification and their spread in values is somewhat larger than the corresponding type of CTD-profiles (Fig. 2a). Although the reason of the difference between CTD and moored T-sensor data is not clear, it may be the result of local (de)focusing of internal wave characteristics. Whilst the overall profiles are consistent throughout, local variations are reflecting partially salinity-compensated intrusions that may or may not be turbulent overturns. The difference in T-sensor and CTD-profiles may also reflect local conditions, the mooring being more in a small seafloor-depression and the CTDs more above a small promontory of a vertical hump, closer to the cliff to the south.

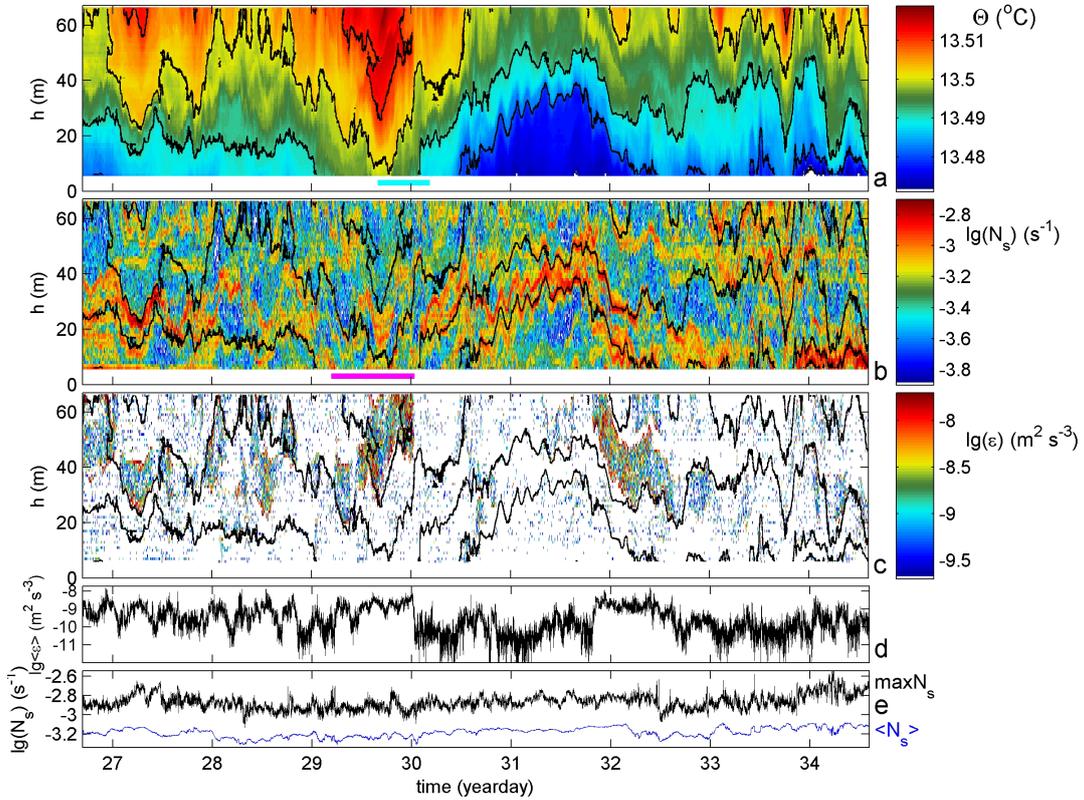

FIG. 5

7.9-Day 60-m-tall, moored T-sensor data and calculated turbulence values. In a.-c., the horizontal axis is at the level of the local seafloor. (a) Conservative Temperature, with black contours every 0.01°C that are repeated in b.,c. for reference. The cyan bar indicates the period of yoyo-CTD. (b) Logarithm of 1-m small-scale buoyancy frequency $N_s$ calculated from reordered profiles of a. The magenta bar indicates the inertial period as in Fig. 4. (c) Logarithm of non-averaged turbulent kinetic energy dissipation rate. White values are below displaying threshold ($<2\times10^{-10}$ m$^2$s$^{-3}$). (d) Logarithm of turbulent kinetic energy dissipation rate averaged over the vertical T-sensor range for each 1-s profile. (e) Logarithm of the maximum (black) and the 60-m vertical mean (blue) of $N_s$ per profile.

Time series of water-flow (current) velocities have speeds of <0.05 m s$^{-1}$, which vary with time over various scales that are mostly <1 day in length, except during the first 3 days when a double-inertial period seems dominant (Fig. 4a,b). Although the acoustic CM-data are relatively noisy, near-inertial periodicities stand out, also in vertical current differences or shear over 215 m vertical distance (Fig. 4c,d). The vertical current component is considerably weaker, with non-noise levels <0.01 m s$^{-1}$, and consistently downward close to the seafloor (Fig. 4e).

The associated 60-m high temperature distribution demonstrates double-inertial and longer period variations, besides near-inertial and shorter period variations (Fig. 5a). Isotherm excursions are several tens of meters high, thereby straining (compressing and depressing) the distance between them, which results in a highly variable pattern of thin and thicker layering of stable vertical density stratification (Fig. 5b). A particular thin (high $N_s > 10^{-3}$ s$^{-1}$; lg($N_s$) > -3) or thick (low $N_s < 3\times10^{-4}$ s$^{-1}$; lg($N_s$) < -3.52) layer never precisely follows an isotherm for longer than an inertial period. Instabilities leading to turbulent overturns and thus dissipation rate of turbulent kinetic energy (Fig. 5c) provide a pattern that generally, but not exclusively associates with the weaker stratification patches. Largest dissipation rates are mostly observed in weak stratification patches in which the temperature distribution seemingly has clear quasi-intrusions of unstable values, e.g., around days 29.7 and 32.1, all for h > 30 m. None of these patches lasts longer than one inertial period and may thus represent turbulent overturns. Closer to the seafloor, such quasi-intrusions are rare, or non-existent, and especially in the first half of the record some localized near-seafloor internal wave breaking is observed for h < 20 m, e.g., thin vertical lines near days 28.05, 29.02 and 30.55, as will become clearer in the magnification around the latter in Section 4.c.

The large dissipation rate patches dominate the vertically averaged values (Fig. 5d), but no consistent correspondence is found anywhere between dissipation rate and (maximum or mean of) small-scale buoyancy frequency (Fig. 5e). A simple parametrization of $\varepsilon \propto N^{-1}$, for 60-m large-scale mean buoyancy frequency N, or $\varepsilon \propto N_s^{-1}$, for 1-m small-scale buoyancy frequency $N_s$, does not seem to dominate these data. Such inconsistency in parametrization of turbulence values with (reordered) stratification was already noted for stronger stratified shear-dominated near-surface waters (Gargett et al. 1981) and for eddy diffusivity determined from moored T-sensor data from Atlantic-Ocean seamount-slope waters (van Haren and Gostiaux 2012).

*b. Mean spectra*

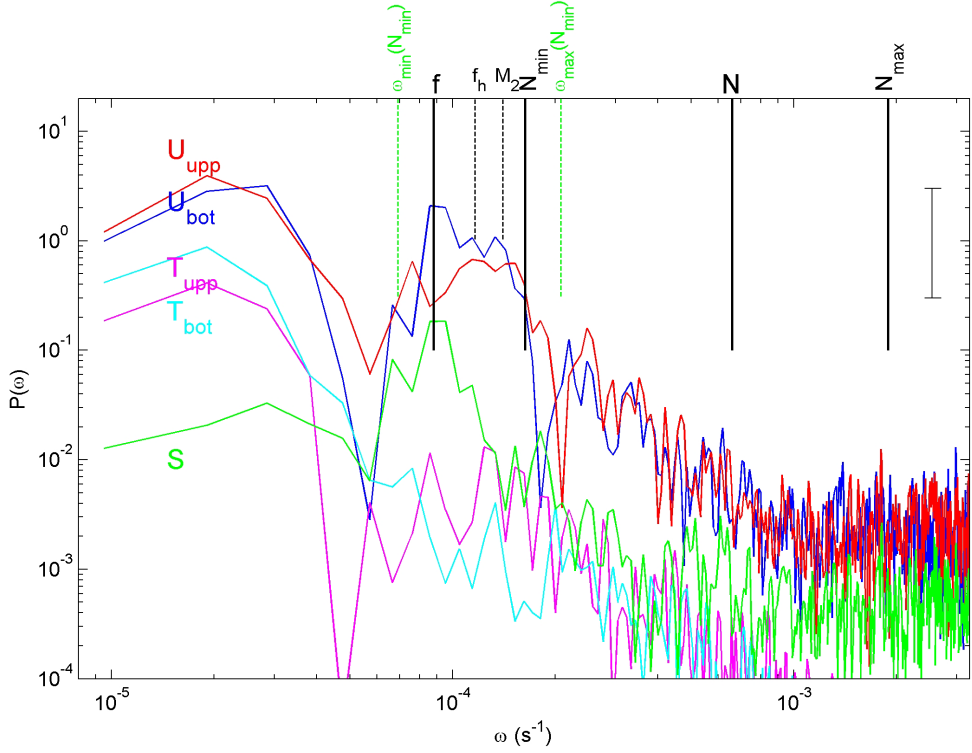

FIG. 6
Weakly smoothed (about 6 degrees of freedom 'dof') frequency ($\omega$) spectra for 7.9-day data from upper (red) and lower (blue) CMs, vertical shear between them (green), and upper (magenta) and lower (cyan) T-sensors. Several frequencies (see text) are indicated with vertical lines, including the non-traditional internal wave band for weak stratification, corresponding with a buoyancy frequency of $N_{min}$, between green-dashed lines. The IGW-bounds would be 15% wider for a buoyancy frequency of $1f_h$, with a minimum IGW-bound at the sub-inertial gap.

Overall, 7.9-day mean CM and T-sensor spectra demonstrate a bulge of relatively large variance at sub-inertial frequencies around $2\times10^{-5}$ s$^{-1}$, or periodicities of about once per 4 days, in both kinetic energy and temperature variance, but not in shear (Fig. 6). While temperature shows otherwise featureless spectra, with a plateau around inertial-tidal frequencies, the CM-data show a sub-peak around these frequencies, more particular in the IGW-band [$\omega_{min}$, $\omega_{max}$]($N_{min}$=2.3f). Throughout the observed vertical range, spectra of a particular variable do not vary significantly with depth, albeit the lower CM shows considerably more inertial energy than higher-up. Most notable is the shear peaking around the local inertial frequency, specifically not at semidiurnal lunar M$_2$. This supports the notion of the importance of near-inertial motions for generating shear-induced turbulence in the ocean interior, due to their short vertical length scales (e.g., LeBlond and Mysak 1978).

To investigate variations in the turbulence range of temperature data as a function of distance from the sloping seafloor, the 7.9-day mean spectra from nine T-sensors are computed (Fig. 7). For focus on the spectral turbulence range, the spectra are scaled with the frequency ($\omega$) slope $\omega^{-5/3}$ of shear-induced inertial subrange (e.g., Tennekes and Lumley 1972).

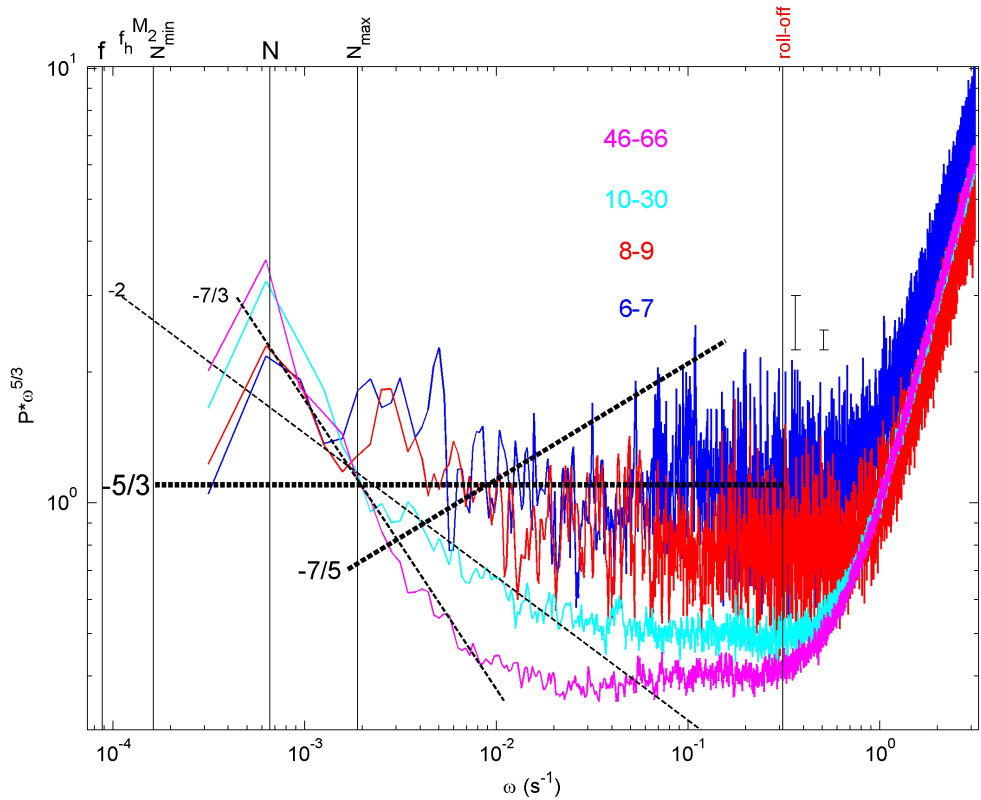

FIG. 7

Temperature variance spectra for 7.9-day data from vertical levels as indicated by their distance h from the seafloor (in meters). The spectra are moderately (about 30 dof; blue and red spectra) and heavily (about 200 dof; magenta and cyan spectra) smoothed and focus on the turbulence range by scaling with the inertial subrange slope $\omega^{-5/3}$. Several frequencies (see text) are indicated with vertical lines, except for horizontal Coriolis parameter f$_h$ and semidiurnal lunar tidal M$_2$. Four spectral slopes are indicated by dashed black lines and by their exponent-value (for unscaled log-log plots; see text).

The plotted reference slopes are indicated by their exponent-value in unscaled plots: $\omega^{-2}$ for spectra of general internal waves outside inertial, tidal (harmonic) and buoyancy frequencies (Garrett and Munk 1972) and also for fine-structure contamination (Phillips 1971; Reid 1971), $\omega^{-5/3}$ for the inertial subrange of turbulence predominantly induced by shear and representing passive scalars (Tennekes and Lumley 1972; Warhaft 2000), and $\omega^{-7/5}$ as an example of significant deviation from inertial subrange for turbulence predominantly induced by buoyancy-driven convection and representing active scalars (Bolgiano 1959; Pawar and Arakeri 2016). The steepest indicated slope of $\omega^{-7/3}$ is a fit to the high-frequency internal wave and low-frequency turbulence part of the buoyancy subrange of the spectrum. It represents intermittency as, e.g., observed in occurrence of sea-surface white-capping (Malila et al. 2022). Local effects of intermittency attributed to coherent structures resulted in a local dip in atmospheric velocity variance at the low end of the inertial subrange (Szilagy et al. 1996).

Vertical black lines indicate frequencies delimiting predominant internal waves at various scales and turbulence. Instrumental noise is basically found only at the far high extreme $\omega > 1$ s$^{-1}$, but roll-off to noise starts at about 0.33 s$^{-1}$. Internal waves are associated with $\omega \in [f, (N_{min}) N (N_{max})]$, of which the range $[f, N]$ is commonly thought to be dominated by undamped linear and unsaturated internal waves and $\omega > N$ to consist of damped nonlinear and saturated internal waves (Weinstock 1978; Munk 1981). The latter have a slope steeper than $\omega^{-7/3}$. Turbulence is found in various identities in the range $\omega \in [N_{max}$, roll-off].

The poorly resolved internal wave band, due to shortness of record and substantial spectral smoothing, shows marginally increasing temperature variance away from the seafloor. For $\omega > N_{max}$, the discrepancy between different heights above the seafloor becomes significantly different to the 95% level, with highest temperature variance nearest to the seafloor and highest discrepancy of about one order of magnitude between near-seafloor (blue/red) and interior (magenta) spectra, at about 0.02 s$^{-1}$.

A relative dip of minimum (scaled) variance is observed around 0.02 s$^{-1}$. The poorly resolved spectral slope exponent (on an unscaled log-log plot) at $\omega > N$ towards this minimum is close to -7/3 for $46 < h < 66$ m and -2 for $10 < h < 30$ m. At $\omega > 0.02$ s$^{-1}$, either the spectral slope exponent follows -7/5 going up in the scaled plot (for h = 6-7 m), remains or becomes approximately horizontal (for h > 8 m). This suggests dominantly shear-induced turbulence as described by a one-order of magnitude resolved inertial subrange for most of the vertical range, except for the lower two T-sensors h ≤ 7 m. These two sensors describe half-an-order of magnitude inertial subrange just before spectral roll-off to noise, but which is preceded by a significantly different slope that approximates the buoyancy subrange slope of convection-turbulence over almost one order of magnitude of frequencies. Over flat topography, dominant convection-turbulence was found in spectra from h < 10 m in the deep Western Mediterranean (van Haren 2023b).

Across the turbulence subranges the temperature variance spreads over twice as much power range as across the noise range. This has been attributed to coherent portions of turbulence (van Haren et al. 2016).

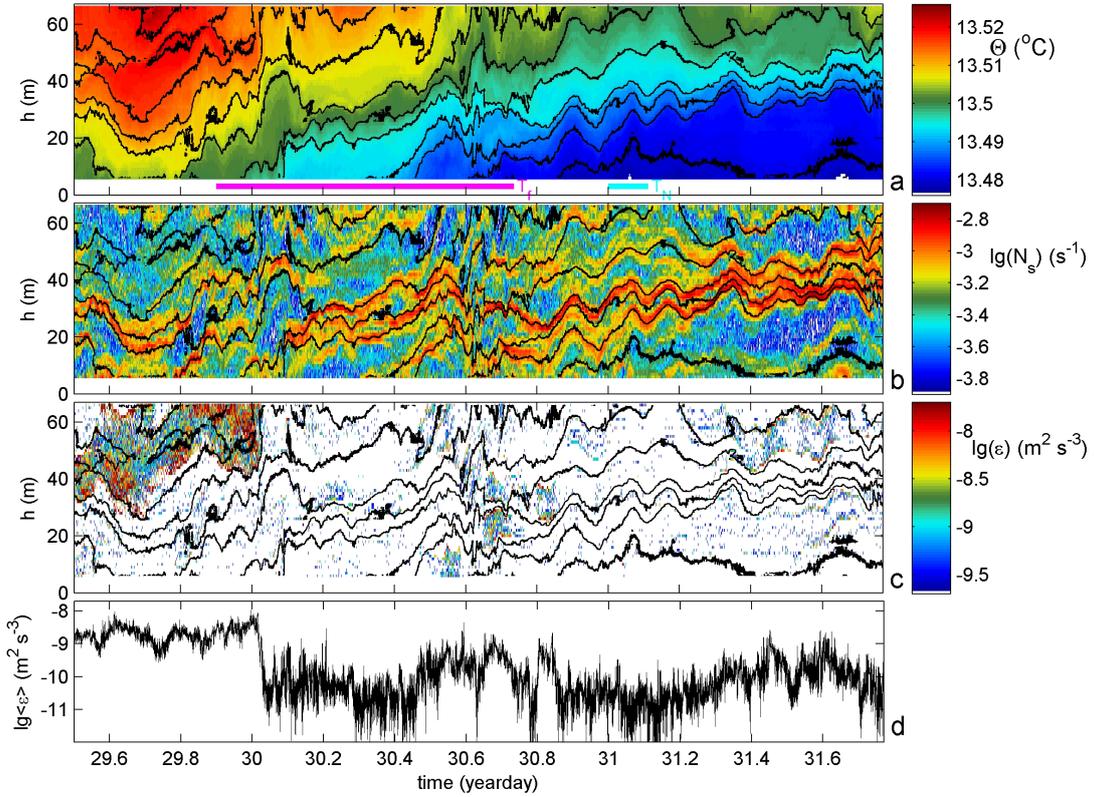

FIG. 8

2.25-Day detail of moored T-sensor observations from warming to cooling phase. In a.-c., black isotherms are drawn every 0.004°C and the horizontal axis is at the level of the local seafloor. (a) Conservative Temperature. The magenta and cyan horizontal bars indicate one inertial and one mean buoyancy period, respectively. (b) Logarithm of small-scale buoyancy frequency $N_s$ calculated from reordered profiles of a. (c) Logarithm of non-averaged turbulent kinetic energy dissipation rate. (d) Logarithm of turbulent kinetic energy dissipation rate averaged over the vertical T-sensor range for each profile.

*c. A glossary of instability details*

Even though the 61-m tall T-sensor string did not resolve all contours of internal wave-dominated excursions over a (sub-)inertial period as several went out of the panel above or below (Fig. 5), some of the peaks in turbulent mixing were found reaching (to within 6 m from) the seafloor, see, e.g., near day 30.6 in the detail of Fig. 8. In this detail, large apparent turbulent overturning is observed in the upper half up to day 30.0. Although the vertically mean turbulence dissipation rate is generally $>3\times10^{-10}$ m$^2$ s$^{-3}$ with variations over barely one order of magnitude, the predominant periodicity of variation is about the buoyancy period (the cyan bar in Fig. 8a may be used as reference in Fig. 8c,d). This is due to short bursts with the generally high values, of <10 m in height. These bursts reflect turbulent overturning and not partially salinity compensated intrusions, while perhaps being embedded in isopycnal transport that provides a generally elevated turbulence dissipation rate value. During the remainder of the period, including the passage of near-seafloor frontal bores of, upslope propagating (Fig. 4a), colder waters, turbulence values are all lower, although occasionally reaching $10^{-9}$ m$^2$ s$^{-3}$ for mean dissipation rate (Fig. 8d). The variations, now over three orders of magnitude, show predominant variability with the mean buoyancy period, as noted before. This periodicity is also evident in the isotherm displacements following the second frontal bore around day 30.6.

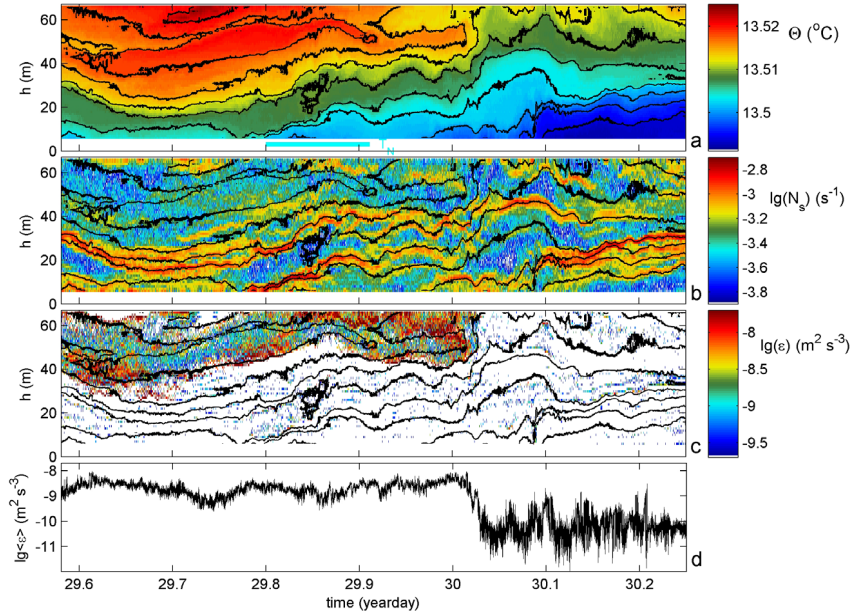

FIG. 9
As Fig. 8, but for 0.7-day focus on the warm-cool transition, with black contours every 0.003°C in a.-c. and different colour range in a. It is noted that $T_{Nmin} \approx 4T_N$ and $T_{Nmax} \approx T_N/3$.

The 61-m and 2.2-d mean turbulent kinetic energy dissipation rate has a value of $6\pm4\times10^{-10}$ m$^2$ s$^{-3}$, while the eddy diffusivity amounts $3\pm2\times10^{-4}$ m$^2$ s$^{-1}$, for mean N = $6.6\pm0.\times10^{-4}$ s$^{-1}$ = 7.5f. These values are comparable with deep near-flat-bottom Pacific Ocean turbulence values (van Haren 2020), about one order of magnitude larger than upper 2000-m open-ocean values (Gregg 1989) and two orders of magnitude smaller than found over steep topography of an underwater seamount (e.g., van Haren and Gostiaux 2012).

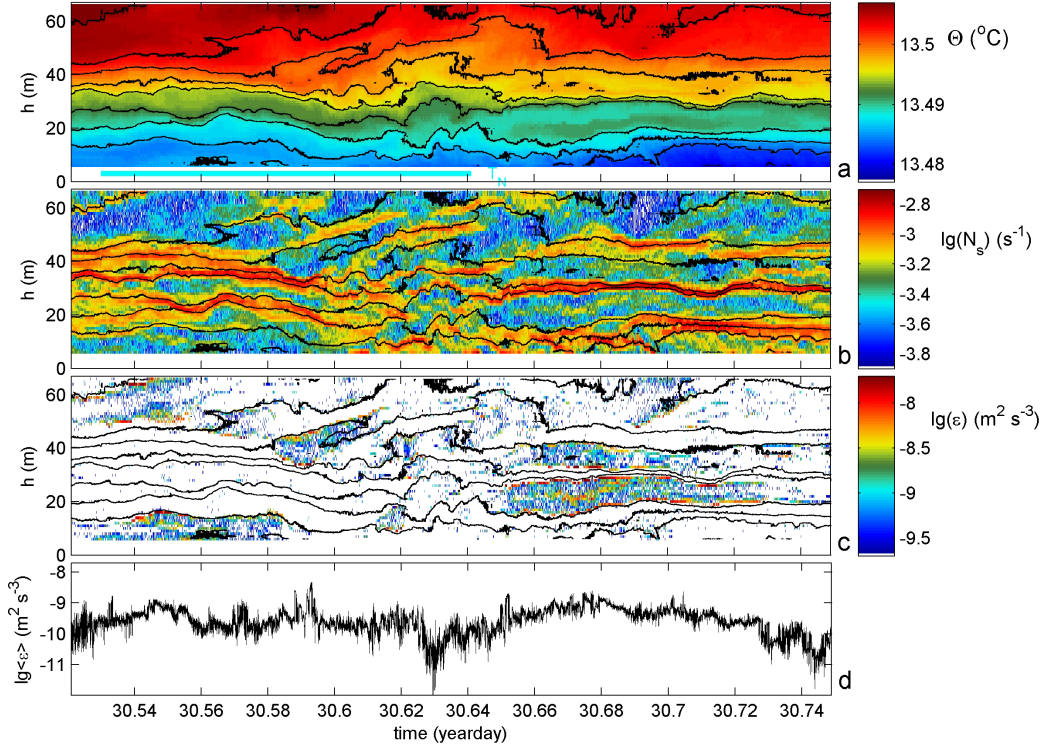

FIG. 10
As Fig. 8, but for 5.5-hour focus on internal wave break-up away from the seafloor, with every 0.003°C black contours in a.-c. and different colour range in a.

Further detailing the apparent intrusion shows that it contributes one order of magnitude larger turbulence dissipation rate than smaller overturns (Fig. 9). Within the large turbulence patch, the short-scale patches of <10 m height and <$T_N$ duration are clearly visible. This patch is further investigated spectrally in Sub-section 4d and in terms of overturn-character in Sub-section 4e, with reference to weaker turbulence patches. In Fig. 9, no difference is observed in isotherm-roughness and weak stratification between large- and weak-turbulence patches. Commonly, turbulent overturns lead to relatively weak stratification via homogenizing and isotherm-straining, resulting in a relatively rough shaping of isotherms. Inspection of Fig. 9 shows that isotherm-straining and -roughness are found throughout the image, which suggests turbulence is active throughout. The considerable differences in value of turbulence are then contributable to the level of (overturn-) instability.

In Fig. 10, the 61-m and 5.5-h mean turbulent kinetic energy dissipation rate has a value of $4\pm3\times10^{-10}$ m$^2$ s$^{-3}$, while the eddy diffusivity amounts $2\pm1\times10^{-4}$ m$^2$ s$^{-1}$, for mean N = $6.7\pm0.\times10^{-4}$ s$^{-1}$ = 7.6f. These values are half an order of magnitude smaller than those for Fig. 9. Despite these lower values, which are mainly attributable to a lack of large upper-level values as in the first half of Fig. 9, the observations in Fig. 10 demonstrate similar albeit less intense elongated turbulent overturns or instabilities. None of these instabilities and associated weakly stratified patches lasts longer than the mean buoyancy period. This corresponds with previous observations made in the abyssal Pacific Ocean (van Haren 2020). One of the patches reaches to within h ≤ 6 m and may touch the seafloor thereby affecting sediment resuspension.

In the center of Fig. 10, around day 30.63, turbulence dissipation rate dips in value, disconnecting turbulent patches before and after. This has resemblance with a similar observation in the abyssal Pacific, where it was attributed to quasi-convective vertical motions. Just before and after the reduction in turbulence however, motions seem to be more like overturning with restratifying motions in between.

*d. Overturn spectra*

The large overturn- or quasi-intrusion-patch of nearly 11-h total duration and about 20-m total thickness at approximately constant height above the seafloor (between 46 < h < 66 m) in Fig. 9 is spectrally investigated, in comparison with data from a reference 20-m thick layer below (10 < h < 30 m). It is observed that the unstable patch (magenta spectrum in Fig. 11) contains (insignificantly) less temperature variance in the internal wave band, but significantly more variance than the reference level (cyan spectrum) at 0.01 < ω < 0.1 s$^{-1}$. Whilst the reference level, nearer to the seafloor, shows approximately constant scaled-spectral level of inertial subrange commensurate dominant shear-turbulence, the intrusion shows an initial increase with maximum slope approaching that of convection-turbulence to ω < 0.08 s$^{-1}$, before a rapid decrease sloping with approximately (unscaled) ω$^{-2}$. Or, differently described, the (unscaled) -7/5-slope exponent of the quasi-intrusion occurs when the reference spectrum -2-slope exponent is found at 0.005 < ω < 0.02 s$^{-1}$, and both spectra are constant (unscaled -5/3-slope exponent) at $0.02 < \omega < 0.09$ s$^{-1}$. The latter inference somewhat makes more sense, as it contains a transition to inertial subrange in both spectra, whilst this subrange is lacking in the former inference for the quasi-intrusion spectrum.

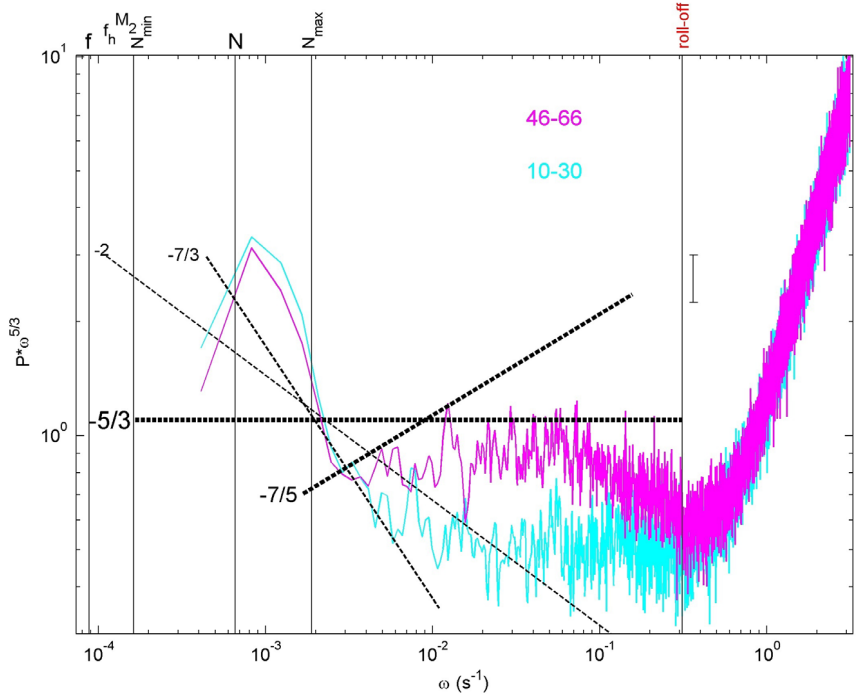

FIG. 11

As Fig. 7, but for 0.44-day spectra distinguishing weakly stratified upper layer quasi-intrusion (magenta spectrum) from non-intrusion data (cyan spectrum) between days 29.8 and 30.02 in Fig. 9. The short data record provided nearly raw unsmoothed spectra for h = 6-7 and 8-9 m, which are left out here, for clarity.

Following the model slopes, and with reference to the turbulence spectra in Fig. 7 notably from h = 6-7 m, the quasi-intrusion spectrum for $46 < h < 66$ m in Fig. 11 describes convection turbulence rather than shear-flow in isopycnal direction. The description of an intrusion should thus be replaced by that of an (elongated) turbulence overturn, with dominant convection-turbulence. This is supported by observing individual profile-overturn shapes below, which were found to describe genuine overturns rather than salinity-compensating intrusions in the abyssal Pacific. (It is noted here that, significantly differently sloping from dominant shear-turbulence in the upper layer, lower-layer spectra of the abyssal Pacific data showed dominant convection turbulence, although it was not named as such in van Haren (2020).

*e. Overturn shapes in displacement profiles*

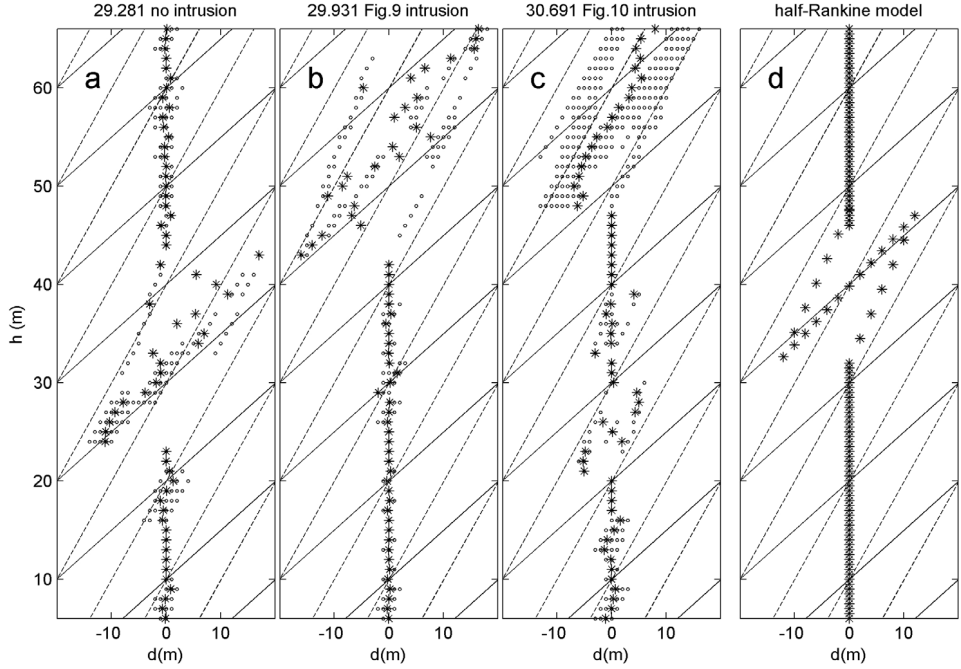

FIG. 12
Vertical shapes after reordering instabilities of 60 s of consecutive T-sensor profiles of displacements (o) and their mean values (*) from times indicated in yeardays as title. The slanted-lines grid consists of two slopes in the d,h-plane: a slope of ½ for solid lines and a slope of 1 for dashed lines. (a) Profiles from data in Fig. 5. (b) Profiles from data in Fig. 9. (c) Profiles from data in Fig. 10. (d) Data from an idealized model of a half-turn Rankine vortex (after van Haren and Gostiaux 2014).

Although the data are less clear than the smoother profiles from the abyssal Pacific (Fig. 7 in van Haren 2020), the displacement-profiles from the Eastern Mediterranean show generally unambiguous results in h/d profiles, see examples in (Fig. 12). While the first profile (Fig. 12a) is from a large turbulent overturn that indeed best resembles a half-turn Rankine vortex (Fig. 12d), smaller overturns are observed in the large one. This affects the mean profile, making it somewhat more erratic. Nevertheless, the typical parallelogram-shape, with steeper upward slopes along the sides than of the long diagonal, are clearly visible. This is also observed in the profiles of Fig. 12b,c, which are obtained across presumed intrusions. Their displacement-profiles appear somewhat more erratic also in the mean (Fig. 12b) and/or with individual profiles being more aligned with $h/d \approx 1$ (Fig. 12c). The latter seems to correspond with buoyancy-driven convection-turbulence with secondary shear-induced overturn instabilities in small parts of a large instability (Appendix B).

The displacement-profiles of Fig. 12 do not show evidence for intrusions, described in Section 2, but rather for turbulent instabilities, either predominantly shear- or buoyancy-driven, thereby confirming the spectral information.

**5. Discussion**

Especially around isolated seamounts, topography-internal wave interactions yield vigorous turbulent mixing that may spread into the surrounding ocean-interior (e.g., Armi 1979; Winters 2015). However, recent observations, using a 1-km high mooring above a steep seamount slope in a tidally dominated East-Atlantic Ocean (van Haren 2023a), demonstrated that, a) internal wave breaking occurred up to a few kilometers from the seamount, b) isopycnal spreading reached the same distance and lasted less than a day in association with the rapid restratification by the back-and-forth sloshing internal waves near the slope, c) near-homogeneous layers showed internal turbulence albeit 100-times weaker than near the slope.

Because of the latter result, mean turbulence profiles increased values with increasing depth by half-one order of magnitude over the lower 70 m (van Haren 2023a). This contrasts with mean turbulence profiles in the Eastern Mediterranean, which decrease with increasing depth by one order of magnitude over the lower 20 m of the T-sensor range from $10^{-9}$-$10^{-10}$ m$^2$ s$^{-3}$ in $<\varepsilon>$ and from $10^{-3}$-$10^{-4}$ m$^2$ s$^{-1}$ in $<K_z>$ with $<N_s>$ weakly increasing with depth, for the data in Fig. 5. The upper turbulence values are only about half an order of magnitude smaller than observed 1 km above the steep East-Atlantic slope.

Considering that for the Eastern Mediterranean mooring the local seafloor slope is about 0.03, observed isopycnal transport around h = 55 m may come horizontally from Δx = 1850±350 m away. This is about the same distance as found away from the steep East-Atlantic seamount-slope under two-three orders of magnitude larger turbulence values near the slope. This apparent similarity of isopycnal extent and turbulence values in considerably different environments suggests little effect from internal tides, from turbulence intensity near the seafloor, and from seafloor slope on local turbulence in layers extending kilometers away from topography. It suggests dominant local (convection-)turbulence generation in the interior which is probably driven by internal wave action and shear or by isopycnal transport from the boundary, but which does not transport boundary-turbulence.

Because stratification rate is one order of magnitude larger (N by a factor of 3) and current speed by a factor of 2 in the East-Atlantic site compared to the Eastern Mediterranean site, it may seem improbable that turbulent values are similar some distance away from topography where most turbulence is generated due to the breaking of internal waves. On the other hand, one could argue that tidally dominated flow speed and stratification rate may not be a good measure for local isopycnal dispersion away from topography. Possibly, stronger, or weaker mean stratification does not prevent straining of isopycnals in the interior, mostly by internal waves, to create thin or thicker layers of variable small-scale stratification and which contain local convection and/or dispersion at the same rate. Likewise, tidal, and inertial waves dominate turbulence generation, which is largest by the breaking of these waves at sloping topography and smaller by their shear in the interior, but they may be less important for isopycnal dispersion that is driven by other flows. Further research from various areas is required on the details of isopycnal redistribution notably near topography to quantify the observed dominance of convection over dispersion from spectral and overturn-shape information.

The observed quasi-intrusions last shorter than the maximum buoyancy period, and thus shorter than the local inertial period with an average duration around the mean buoyancy period. Their overturns compare well with turbulent overturn-models while yoyo-CTD profiles show a consistent temperature-density relationship also in unstable layers. This supports the notion that unstable layering is likely to generate turbulence locally (as visible in online video material of Winters 2015), rather than transporting (weaker) turbulence generated at the seafloor along the isopycnals. As a consequence, establishment of horizontal diffusivity has little meaning for such quasi-intrusions, which do not show dominant horizontal dispersal as, e.g., in mesoscale eddies.

The spectral information shows a small tendency for convection-turbulence in weakly stratified layers, as well as more generally closest to the seafloor, in an otherwise dominant shear-turbulence environment in which internal-wave breaking associated with frontal bores seldom occurred. The moored T-sensor observations are consistent with Arctic-Ocean observations, where freely propagating lunar internal tides cannot exist, of diurnally varying turbulent patches that are also mainly driven by convection in the interior made using shipborne (shear-probe) microstructure profiler, which, however, provided about one-order of magnitude larger turbulence values (Padman and Dillon 1991).

Thus, on relatively short internal wave-scales the variation in stratification renders important transitions, with the minimum buoyancy frequency determining the large-scale IGW-energy bounds and the timescale of multiple overturn duration, with the mean buoyancy frequency setting the typical timescale of largest individual overturns, and with (twice) the maximum buoyancy frequency leading to a transition between shear- and convection turbulence.

On larger scales, the spectral peak in shear around the local inertial frequency was expected following the knowledge of short vertical length scales of near-inertial motions (e.g., LeBlond and Mysak 1978). To somewhat lesser extent, the enhancement of kinetic energy in the limited IGW-band was expected for weakly stratified waters of minimum stratification. Unknown is the poorly resolved enhanced energy around $\omega = 2\text{-}3\times10^{-5}$ s$^{-1}$, which is outside the IGW-band and may be associated with (sloping) boundary flows. Such flows may develop (sub-)mesoscale activity, but their presence was outside the scope of the short-duration mooring. Their potential interaction with the IGW, possibly via the topography and potentially developing dispersion, may be worth a future study of longer duration moorings.

*Acknowledgements.*

I thank the captain and crew of the R/V Meteor and engineers from NIOZ-MRF for their assistance during instruments and mooring preparation, deployment, and recovery. NIOZ temperature sensors have been funded in part by NWO, the Netherlands organization for the advancement of science.

*Data availability statement.*

Current meter and raw SeaBird-911 CTD data supporting the results of this study are available in database https://data.mendeley.com/datasets/6td5dxf6bj/1. Moored T-sensor data require extraction from a raw-data base and are available from the author upon reasonable request.

**Appendix A. Computing turbulence from moored T-sensor data**

Commonly, shipborne CTD-profiles are made as close as possible near a mooring containing high-resolution T-sensors. This is to achieve local information on vertical temperature profiles via recently calibrated instrumentation and on the temperature-density relationship, such as (1). During the present cruise, the nearest CTD-profiles were obtained some 3.5 km south of the T-sensor mooring, where the water depth was about the same as at the mooring site, but with unknown variations in small-scale topography.

Nonetheless, given the relationship (1), the turbulence parameters ε and $K_z$ can be calculated for moored T-sensor data using temperature as a proxy for density variations following Thorpe's (1977) method of reordering unstable vertical overturns. Thus, every 1 s the 60-m high potential density (Conservative Temperature) profile $σ_{1.2}(z)$ is reordered, which may contain unstable inversions, into a stable monotonic profile $σ_{1.2}(z_s)$ without inversions.

After comparing observed and reordered profiles, displacements d = min(|z-$z_s$|)·sgn(z-$z_s$) are calculated necessary to generate the reordered stable profile. Test-thresholds are applied to disregard apparent displacements associated with instrumental noise and post-calibration errors (Galbraith and Kelley 1996). Such a test-threshold is very low $<5×10^{-4}$ °C for originally sampled non-spiked moored NIOZ T-sensor data. Otherwise, identical methodology is used as proposed for CTD-data in (Thorpe 1977). It includes application of a constant mixing efficiency of 0.2 for shear-dominated turbulence in stratified waters (Osborn 1980; Oakey 1982), an Ozmidov $L_O = (ε/N^3)^{0.5}$ (Dougherty 1961; Ozmidov 1965) -- root-mean-square (rms) overturn scale $d_{rms} = (1/nΣd^2)^{0.5} = L_T$ ratio of $L_O/L_T = 0.8$ (Dillon 1982) over many-n samples, and the computation of local small-scale buoyancy frequency $N_s(t,z)$ from the reordered stable density (temperature) profiles (Thorpe 1977; Dillon 1982).

Although individual values of mixing efficiency vary over at least one order of magnitude (Oakey 1982; Dillon 1982), moored T-sensor data allow for sufficient averaging to warrant the use of constant mixing efficiency in the computation of turbulence values. Every second a vertical profile is made, so that averaging over the mean buoyancy or over the inertial period or longer provides sufficient statistical means for an average value of mixing efficiency, like provided for considerably less profiles by, e.g., Oakey (1982) and Dillon (1982). When sufficient averaging over time and/or depth is applied, various types, conditions and stages of turbulence are included, each with different efficiency (Mater et al. 2015; Cyr and van Haren 2016). Large-scale N is computed as rms($N_s$), an average over the entire vertical range of T-sensors. This is a linear operation as $N^2 \propto d\sigma_{1.2}/dz$. Then, using d rather than $d_{rms}$ as explained below,

$\varepsilon = 0.64d^2N_s^3$, (A1)

$K_z = 0.2\varepsilon N_s^{-2}$. (A2)

In (A1), and thereby (A2), individual d replace their rms-value across a single overturn as originally proposed by Thorpe (1977). Also, $N_s$ is used instead of N. In the past some tests were done using mean N, for different data sets. Although the results were generally within a factor of 2-3 approximately, it was decided to use to more appropriate local $N_s$, as employed for the present data. The reason is that individual overturns cannot easily be distinguished, because overturns are found at various scales with small ones overprinting larger overturns, as one expects from turbulence. This procedure provides high-resolution time-vertical images of turbulence values, for qualitative studies. Subsequently for quantitative studies, 'mean' turbulence values are calculated by arithmetic averaging over the vertical <...> or over time [..], or both, which is possible using moored high-resolution T-sensors. This ensures the sufficient averaging required to use the above mean constant values (e.g., Osborn 1980; Oakey 1982; Mater et al. 2015; Gregg et al. 2018).

Using similar moored T-sensor data from GMS, van Haren and Gostiaux (2012) found turbulence values to within a factor of three like those inferred from ship-borne CTD/Lowered-ADCP profiler data using a shear/strain parameterisation near the seafloor. Their values compare well with microstructure profiler-based estimates in similar sloping topography areas by Klymak et al. (2008). Comparison between calculated turbulence values using shear measurements and using overturning scales with $L_O/d_{rms} = 0.8$ from areas with such mixtures of turbulence development above sloping topography led to consistent results (Nash et al. 2007), after suitable averaging over the buoyancy scales. Thus, from the argumentation above and the reasoning in Mater et al. (2015), internal wave breaking is unlikely to bias turbulence values computed from overturning scales by more than a factor of two to three, provided some suitable time-space averaging is done instead of considering single profiles. This is within the range of error, also of general ocean turbulence measurements using shipborne microstructure profilers (Oakey 1982).

**Appendix B. Overturn displacement-shapes from geothermal convective heating**

A (half-turn) Rankine-vortex seems a reasonable model for shear-induced turbulent overturning, providing displacement-profiles that resemble those observed in well-stratified ocean data (van Haren and Gostiaux 2014). However, this vortex-model may not be suitable for buoyancy-driven convection-turbulence. This is because convection is generally modelled as up- and down-going tubes (or plumes) of motions (e.g., Marshall and Schott 1999), although shear may develop secondary overturn instabilities along the sides of such plumes (Li and Li 2006).

As a rare example of directly observed deep-sea convection, displacement-profiles are calculated for high-resolution T-sensor data from a mooring in the deep Western Mediterranean in extremely weakly stratified waters over general geothermal heating (Fig. B1). The displacement profiles appear rather chaotic, although generally aligned with h/d ≈ 1. Only in small parts, best noticeable in the mean (star) values, interior-overturn alignment is observed, e.g., around h = 7, 40, 85 and 100 m.

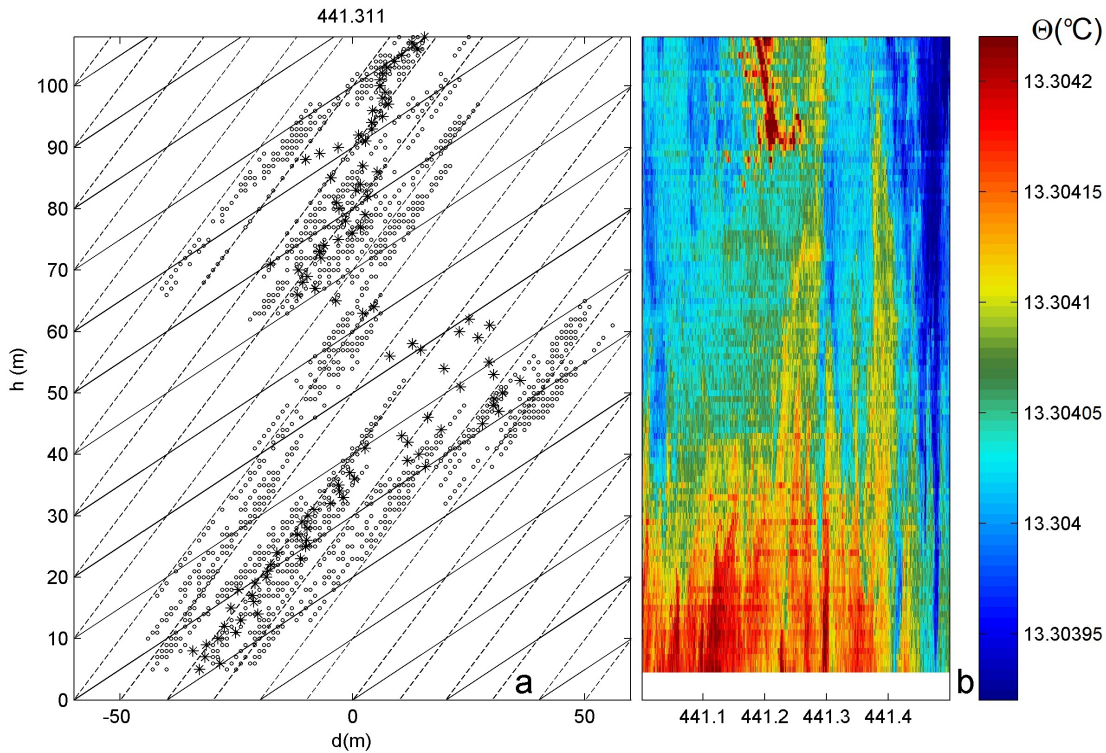

FIG. B1
Displacement-profiles determination for directly observed geothermal convective heating using T-sensors moored between 5 < h < 108 m above a flat 2480-m deep Western Mediterranean seafloor (van Haren 2023b). (a) As Fig. 12, but for displacement-profiles from data in b. (b) Time-depth distribution of Conservative Temperature.